\documentclass[12pt]{article}
\usepackage[utf8]{inputenc}
\usepackage{graphicx} 
\usepackage{dcolumn} 
\usepackage{bm} 
\usepackage{amsfonts}
\usepackage{amsthm}
\usepackage{amsmath}
\usepackage{amssymb}
\usepackage{epsfig}
\usepackage{color}
\usepackage{textcomp}
\usepackage{hyperref}
\usepackage{titlesec}
\usepackage{slashed}
\usepackage{caption}
\usepackage{subcaption}
\usepackage{siunitx}
\usepackage{booktabs}
\usepackage{multirow}
\usepackage{cite}
\usepackage{placeins}
\usepackage{comment}
\usepackage{braket}
\usepackage{authblk}

\let\a=\alpha \let\b=\beta  \let\g=\gamma   \let\e=\varepsilon
      \let\k=\kappa \let\l=\lambda
\let\m=\mu    \let\n=\nu             \let\r=\rho
 \let\t=\tau    
   
 \let\D=\Delta   
     \let\F=\Phi

\def\\{\hfill\break} \let\==\equiv

\let\0=\noindent

\let\dpr=\partial  
\def\nn{\nonumber}

\def\qed{\hfill\raise1pt\hbox{\vrule height5pt width5pt depth0pt}}

\def\be{\begin{equation}}
\def\ee{\end{equation}}
\def\bea{\begin{eqnarray}}\def\eea{\end{eqnarray}}
\def\bs{\boldsymbol}
\def\mc{\mathcal}

\begin{document}
\title{Does $E=mc^2$ Require Relativity?}
\author{Tony Rothman \thanks{tonyrothman@gmail.com}}
\affil{\small New York University, Department of Applied Physics (retired)}
\date{\today}

\markright{$E=mc^2$...}

\maketitle

\begin{abstract}

It is universally believed that with his 1905 paper ``Does the inertia of a body depend on its energy content?" Einstein first demonstrated the equivalence of mass and energy by making use of his  special theory of relativity.  In the final step of that paper, however, Einstein equates the kinetic energy of a body to its Newtonian value, indicating that his result is at best a low-velocity approximation. Today, several characters debate whether a mid-nineteenth century physicist, employing only physics available at the time, could plausibly arrive at the celebrated result.  In other words, is Einsteinian relativity necessary to derive ${\mc E}=mc^2$?

\end{abstract}

\section{At sunrise...}
\setcounter{equation}{0}\label{intro}
\baselineskip 8mm
\begin{flushleft}
{\it Salviati}: I am pleased to know, Simplicio, that you have heard of ${\mc E}=mc^2$.  I was becoming concerned that you were entirely ignorant of twentieth-century physics.\\
\vspace{3mm}
{\it Simplicio}: That I am ignorant, Salviati, I admit freely, and of this equation in particular.  When I try to understand it, I find myself all at sea.\\
\vspace{3mm}
{\it Salviati}: Truly?  Pray tell, what is there to understand?  In 1905, employing his theory of relativity, Albert Einstein proved that any material body contains an amount of energy equal to its mass multiplied by the square of the speed of light.  That, I assure you,  is the beginning and end of the story.\\
\vspace{3mm}
{\it Simplicio}: I must be blind.  When I read his paper I fail to see any relativity whatsoever,  at least Einstein's.\\
\vspace{3mm}
{\it Salviati}: How can that be possible?  Are you certain you have read it?  It is entirely based on his special theory, which he created only   months earlier.\\
\vspace{3mm}
{\it Simplicio}: Unless it was a  dream, I have surely read it and find myself scratching my head.  If Einstein began the paper with relativity, then he seems to have lost it by the finale.  I believe a person cleverer than I could reach the same conclusion using only the physics of our master Galileo, perhaps Maxwell's...\\
\vspace{3mm}
{\it Salviati}: Simplicio, as everyone knows, Maxwell's theory is entirely compatible with Einsteinian relativity, while  Galileo's is not.  In your deliberations you must have been unwittingly employing Einstein all along, thus flagrantly contradicting yourself.\\
\vspace{3mm}
{\it Simplicio}: Forgive me, Salviati, I misspoke.  Allow me to suggest that one can reach Einstein's conclusion with only the physics known to Galileo and any practitioner of electromagnetism in the decades following Faraday and Amp\`ere, let us say until about the time Maxwell published his theory in 1865.\\
\vspace{3mm}
{\it Salviati}: Surely that is impossible.  You have blundered into some grave misunderstanding.  However, I propose that we postpone this discussion until after tea.  Ah, here it is...!\\
\end{flushleft}

\vspace{8 mm}

In suggesting that the debate be postponed, Salviati is revealing a certain canniness, for he understands the howls of protest that  any excursion into the origins of relativity elicits   from fellow natural  philosophers.  In the  century since its debut, special relativity has unequivocally become the common property of all philosophers and physicists, and few have been known to alter their positions in any exchange regarding its interpretation.  We nevertheless tend to agree with Bellac and L\'evy-Leblond \cite{BLB73} in that the real drawback in any discussion about the fundamentals of relativity lies in   ``the lack of rigor and vagueness" which characterizes the debate. To fulfill the task our protagonists have framed,  we must thus make more  precise exactly what  poor Simplicio means  by ``Galilean" or ``perhaps-Maxwellian" physics.  Let us first, however, briefly review the historical and scientific misconceptions surrounding the world's most famous equation, which will also  help define the field of combat.

With his  1905 paper ``Does the inertia of a body depend on its energy content?" \cite{Ein05b}, Einstein  became the first person to consider via a thought experiment the {\it conversion} of matter to energy, but he was far from the first to consider an {\it equivalence} of mass and energy. Beginning with Thomson in 1881 \cite{Thom81}, a number of attempts had already been made to relate the mass of a charged object to the energy of its  electromagnetic field. Thomson argued that the presence of the magnetic field produced by a moving, charged sphere  would impede the sphere's motion,  resulting in  an apparent mass increase amounting to $(8/15) {\mc E}/c^2$ for electric field energy ${\mc E}$. Heaviside \cite{Heavi89} later corrected Thomson's result to show that the ``induced" mass (in this case the full mass) should be  $m = (4/3){\mc E}/c^2$, a result also produced by Abraham's classical model of the electron \cite{Ab03}. 

In 1900, Poincar\'e  \cite{Poin00}  argued that an electromagnetic field carried momentum and behaved like a ``{\it fluide fictif}"  such that the field energy ${\mc E}=mc^2$ (cf. \S 6 below). In a trilogy of papers from 1904 and early 1905, Fritz Hasen\"{o}hrl \cite{H1,H2,H3} considered blackbody radiation contained in a perfectly reflecting moving cylinder, and concluded that the radiation behaved as if it had an effective mass $m = (4/3){\mc E}/c^2$, where ${\mc E}$ now became the energy of the blackbody radiation. None of these investigators considered the actual conversion of matter into energy, which  was left to  Einstein and which was probably his greatest contribution to the problem.\footnote{For further historical discussion on these matters see Jammer \cite{Jammer51}, Fadner\cite{Fadner88} and Rothman \cite{Roth21a}.  For technical treatments concerning how the ``4/3 problem" can be resolved in the context of special relativity  see Rothman and Boughn \cite{BR11, Roth21, Boughn13}. The 4/3 is a geometric factor that arises because one is performing a spherical integration over the field/radiation, which is assumed to be isotropic.}

For the present debate, we, along with Simplicio and Salviati, are  interested in discovering what assumptions are necessary to provide a ``plausible" demonstration of ${\mc E}=mc^2$ and whether those assumptions require Einsteinian relativity or merely Galilean relativity.  The word ``plausible" alone is highly subjective, and partly for that reason  the  task turns out not to be  straightforward.   At this juncture, in fact, we are abruptly faced with the first objection,  suggested by Salviati:  Didn't Einstein himself answer the question  in 1905?

Simplicio has indicated ``no." He is troubled not only by the deficiencies of Einstein's original  paper but by  common misconceptions surrounding proofs of ${\mc E}=mc^2$, as well as by misunderstandings of the very notion of proof.  Many physicists, for instance, are under the impression that ${\mc E}=mc^2$ can be established by employing the four-vector formalism of special relativity. An early draft of  Wikipedia's page on mass-energy equivalence in fact offered exactly such a ``derivation."  Four-vectors, however, are \emph {defined} in order to be consistent with ${\mc E}=mc^2$; consequently any argument based on them to prove the relationship is circular. (When an anonymous reader pointed this out to the Wikipedia editors the ``derivation" was removed.)

A universal, assumption-free proof of ${\mc E}=mc^2$ is no more attainable than a universal proof of conservation of energy or momentum, and the very idea that all physics can be derived from a master Lagrangian without  experimental input must be doomed to failure.  For that reason, all demonstrations of mass-energy equivalence rely on specific assumptions and approximations.  The closest thing that exists to a general proof of ${\mc E}=mc^2$ is the Laue-Klein theorem \cite{Laue11, Klein18, Ohan12} of 1911 and 1918, which in essence states that {\it if} ${\mc E}=mc^2$ holds for a point mass, then it also holds for an extended closed system, under specified boundary conditions.  If radiation can escape to infinity, for example, the boundary conditions are evaded.

 Einstein  was aware of the inadequacies of his 1905 article and attempted to correct them in six further papers, but as Ohanian argues \cite{Ohan09},  none is free of errors and inconsistencies.  Physicists who have actually read the 1905 paper know that the dubious step is the final one, in which Einstein relies on the Newtonian value for the kinetic energy.  Hence his  result cannot be regarded as fully relativistic and Simplicio may be forgiven for raising the natural question, Can one arrive at ${\mc E}=mc^2$ in a consistent and plausible manner using only Galilean mechanics and ``perhaps Maxwellian'' electrodynamics?

 We will find that the answer to this question is fairly subtle, requiring us to   distinguish not only between Galilean relativity (GR) and Einsteinian relativity (ER), but various approximations within each.  Our approximations will always be based on the quantity $\b \equiv v/c$ for velocity $v$ and speed of light $c$, and  ``first order,"  ``second order," etc. always refers to powers of $\b$.

 In \S\S 2-3 we review Einstein's 1905 derivation, which was based on energy considerations and introduce a well-known variant based on momentum conservation.  In \S 4 we discuss Galilean  transformations of the electromagnetic field and ask whether the relationship $p = {\mc E}/c$ holds in the Galilean case.  In \S 5 we scrutinize the validity of these transformations in light of the arguments of Bellac and L\'evy-Leblond \cite{BLB73}.  In \S 6 we apply our considerations to Einstein's 1906 derivation of ${\mc E}=mc^2$, and as the sun sets in \S 7 we  contemplate the possibility of deriving ${\cal E}=mc^2$ in other theories and further ponder the meaning of  the results...
\begin{flushleft}
{\it Salviati}: I like nothing more than a good debate.\\
\vspace{3mm}
{\it Simplicio}: I like tea...
\end{flushleft}

\section{Einstein's 1905 Derivation}
\setcounter{equation}{0}\label{Einstein}

\begin{figure}[htb]
\vbox{\hfil\scalebox{.6}
{\includegraphics{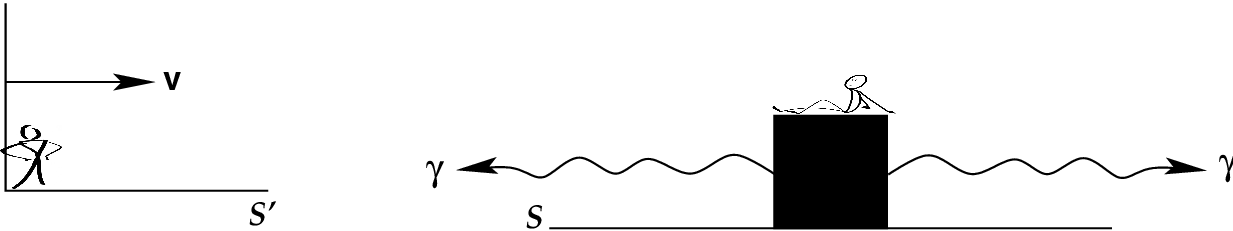}}\hfil}
{\caption{\footnotesize{Einstein's thought experiment. A mass $m$,  at rest in the frame $\cal S$, emits two equal and opposite bursts of electromagnetic radiation.  Observers in a frame, $\cal S'$, moving at a velocity $\bf v$ with respect to $\cal S$ will see the left-moving burst Doppler blueshifted and the right-moving burst redshifted.\label{block1}}}}
\end{figure}

In his 1905 paper, Einstein effectively considers a point mass that  emits two equal bursts of electromagnetic radiation in opposite directions (Figure \ref{block1}).    In a lab frame $\cal S$ where the mass is initially at rest, it  clearly  remains at rest.  Einstein assumes that the radiation carries away an energy ${\mc E}_\g$ such that
\be
{\mc E}_i = {\mc E}_f + \frac1{2} {\mc E}_\g + \frac1{2} {\mc E}_\g, \label{Eno}
\ee
where ${\mc E}_i$ is the initial energy of the mass and ${\mc E}_f$ is its final energy.   As in Figure \ref{block1}, observers in  a rocket frame $\cal S'$ moving at velocity $+v$ with respect to the lab, will see the left- and right-moving radiation  Doppler blueshifted and redshifted, respectively.  Einstein employs the relativistic Doppler shift, which he had previously derived in his ``On the Electrodynamics of Moving Bodies"\cite{Ein05a}, to get
\bea
{\mc E}'_i &=& {\mc E}'_f + \frac1{2} {\mc E}_\g \left(\frac{1- \b}{\sqrt{ 1-\b^2}}\right)+ \frac1{2} {\mc E}_\g \left(\frac{1+ \b}{\sqrt{1-\b^2}}\right)\nn\\
&=& {\mc E}'_f + \frac{{\mc E}_\g }{\sqrt{1-\b^2}}.  \label{En'o}
\eea

Subtracting equation (\ref{Eno}) from equation (\ref{En'o}), Einstein argues that ${\mc E}'-{\mc E}$ must be a kinetic energy, i.e., ${\mc E}_i'-{\mc E}_i = (1/2)m_i v^2$ and ${\mc E}_f'-{\mc E}_f = (1/2)m_f v^2$.  Then   
\be
\frac{1}{2}\D mv^2 = {\mc E}_\g \left(\frac1{\sqrt{1-\b^2}} - 1\right),\label{En1}
\ee
where $\D m \equiv m_i - m_f$.  But  to evaluate this expression he expands the square root as $(1-\b^2)^{-1/2} \approx 1 + (1/2)\b^2 + (3/8)\b^4 + ...$.  Ignoring terms higher than $\mc{O}(\b^2)$ leads immediately to
\be
{\mc E} = \D m c^2.
\ee

We see that although Einstein has begun with the relativistic Doppler effect, in the final step he has equated the $\b^2$ term to the Newtonian kinetic energy. Consequently, one must regard his derivation as only a low-velocity approximation, an observation made as early as 1907 by Planck \cite{Planck07}.  At the time, of course, Einstein did not know the correct relativistic expression for kinetic energy, which was only introduced by Minkowski three years later \cite{Mink08}. For the calculation to be consistent, any  energy on the left hand side of (\ref{En1}) must be proportional to $(\g - 1)$,  where the Lorentz factor $\g \equiv \frac1{\sqrt{1 - \b^2}}$; otherwise one is equating unequal powers of $\b$. With perfect hindsight, one can say that the left side in (\ref{En1}) should be  exactly $\D mc^2 (\g - 1)$. However, that expression is precisely the definition of relativistic kinetic energy and so the proof becomes tautological.

 Einstein's calculation evidently gives the ``correct" answer because Newtonian kinetic energy, $(1/2)mv^2$, is already of order  $v^2$, where relativistic effects typically enter.  The same thought experiment employing the classical Doppler shift (that is, ignoring the $\b^2$ terms above) leads to an ``incorrect" result ($\D mc^2 = 0$). Nevertheless, it is not immediately apparent to what extent Einstein's equation contains relativistic effects.  At the least he has ignored corrections of order $\b^4$ on both sides.  We are therefore entitled to ask with Simplicio whether the derivation might be accomplished beginning from Newtonian physics and halting at order $\b$, which would ensure that no relativistic corrections are introduced.  Such an approach suggests considering momentum, $mv$, rather than kinetic energy $(1/2)mv^2$.
\begin{flushleft}
{\it Simplicio }: Oh, that sounds entirely reasonable to me.\\
\vspace{3mm}
{\it Salviati}: Simplicio, you are an optimist.
\end{flushleft}

\section{$\mc{E} = mc^2$ by momentum conservation}
\setcounter{equation}{0}\label{Momentum}

 Proofs of ${\mc E}=mc^2$ based on conservation of momentum have in fact been provided many years ago by Steck and Rioux \cite{SR83} and Rohrlich \cite{Rohr90}.   The thought experiment is substantially the same as above: At time $t=0$ a mass $m$  at rest in the lab emits equal bursts of radiation in opposite directions.  Because in the lab frame $\cal S$, the mass is initially at rest and remains so, momentum conservation requires that both initial  and  final momentum be zero.  In the rocket frame $\cal S'$ the block has initial velocity $\bf v'$ and, since it remains at rest in the lab, it must remain moving at $\bf v'$ in $\cal S'$.  Momentum conservation then requires that the mass of the block  has changed:
\be
m_f {\bf v'} = m_i {\bf v'} - {\bf p'_{\g +} -  p'_{\g -}},\label{mom}
\ee
where ${\bf p}'_{\g\pm}$ represents the momentum lost to the radiation in the right and left directions, respectively. 

 Both Steck and Rioux and Rohrlich now enlist a photon picture in which ${\mc E}_\g=h\n$ and $p_\g=h\n/c$.  Then, restricting ourselves to one dimension with $\cal S'$ moving at, say, $+ v$ ($v' = -v$), the classical Doppler shift gives 
\be
m_f v' = m_i v' - \frac{{\mc E}_\g}{2c}(1 - \b) + \frac{{\mc E}_\g}{2c}(1 + \b),  \label{mom1}
\ee
and it follows at once that ${\mc E} = \D mc^2$.

This derivation has assumed momentum conservation with classical expressions for the momentum, the classical Doppler shift,  a constant speed of light and the quantum-mechanical expressions for photon energy and momentum, which imply $p_\g = {\mc E}_\g/c$.  
\begin{flushleft}
{\it Simplicio }: I am scratching my head.  Which of these factors are relativistic?\\
\end{flushleft}

A reasonable question.  The classical Doppler shift tacitly assumes that light  is traveling at an absolute velocity $c_o$ through a stationary rest frame---the ``luminiferous aether"---and that the observer and/or source moves at an absolute velocity $\bf v$ with respect to that frame.  To order $v/c_o$  the classical Doppler shift formulas give the above result.  No ER is involved.   However, Simplicio might observe that in an aether theory, the speed of light should not be considered constant, but obey the rules of Galilean velocity addition as any material object (\S 4, below).  To an observer in $\cal S'$, then, the speed of light should be $c_o\pm v$. Salviati, however, points out that substituting  $c_o \pm v$ into (\ref{mom1}) leaves the result unchanged to order $v/c$, so we would again agree that this step remains within  the realm of classical physics, or GR. 

The most serious assumption in the above proof of course lies in using the quantum mechanical relationships  ${\mc E}_\g=h\n$ and $p_\g={\mc E}_\g/c$.  Poynting's theorem (\S 6),  which follows from Maxwell's equations, assures us that $p ={\mc E}/c$ is true for Maxwellian electromagnetic waves.  Compton \cite{Compton23} established that the same should be true for photons, while   Einstein himself had already argued \cite{Ein16} that the relationship must hold for photons in order to obtain the spectrum of blackbody radiation.  Because blackbody radiation is inherently an electromagnetic phenomenon, $p_\g ={\mc E}_\g/c$  for photons must  be compatible with Maxwellian electromagnetism.  Crucially, in turn, Maxwell's equations are ``Lorentz invariant," i.e. compatible with ER (see \S 5), and thus we regard the introduction of photons as implicitly assuming ER.

One might counter that in setting $p = mv$, rather than the relativistic $p=\g mv$, and in using the classical Doppler effect, we have introduced a m\'elange of ER and GR assumptions.  Evidently this is  true; however, once again, going to  fully relativistic expressions for momentum and the Doppler shift gives a tautological result.
\begin{flushleft}
{\it Simplicio }: For me the essential question  has become whether the relationship for light
\be
p = \frac{{\mc E}}{c} \label{pEc}
\ee
implicitly assumes  Clerk Maxwell's electrodynamics or is more general.  If it can be obtained without Maxwellian physics, can one  not maintain that a conservation of momentum proof such as that above,  does not require ER?\\
\vspace{3mm}
{\it Salviati}: Hmm, I perceive that this may not be a trivial issue.
\end{flushleft}

 Yes, Salviati, this is far from a trivial issue.  Electromagnetic waves are transverse and Simplicio may be puzzled by the fact that a purely transverse wave can carry any momentum whatsoever in the direction of propagation.  According to Peskin \cite{Pesk10}, $p = {\mc E}/c$ holds not only for light, but for (longitudinal) sound waves, water waves, and waves on a nonlinear vibrating string, but not for a linear transversely vibrating string.  He therefore reasonably doubts that a universal proof of $p = {\mc E}/c$ exists.  That conclusion is apparently supported by Roland and Pask \cite{RP99} who point out numerous exceptions to the $p={\mc E}/c$ rule in the literature on elastic strings. 

 Because today we are not considering strings but the conversion of matter to energy we  shall  restrict our attention to electromagnetism.  A central question for the remainder of the debate  becomes Simplicio's: Can one produce a valid, non-ER proof of ${\mc E}=mc^2$ assuming \ref{pEc}? By ``valid" we mean correct to order $\b$.  At that level of approximation, one will not  know whether the result is  correct to higher orders, but that is  much the point.  In order to intelligently discuss this question, let us first examine the Galilean transformations of the electric and magnetic fields.
\begin{flushleft}
{\it Salviati}: Allow me  at this moment to propose a wager: your choice of wine says that such a nineteenth-century proof cannot truly be had.\\
\vspace{3mm}
{\it Simplicio}: Very well, my friend, Galileo has  never been known to refuse a bottle.\\
\end{flushleft}

\section{Galileo is Summoned}
\setcounter{equation}{0}\label{Galileo}

Our immediate task is to examine the transformation of the electric and magnetic fields under the rules of Galilean relativity in order to derive  Doppler-shift formulas analogous to those employed above.  The fundamental assumption of GR, which ER adopts, is that the laws of physics are the same in all inertial, or nonaccelerating, reference frames.  As a consequence, two identical experiments performed in two different inertial frames must give identical results.  The most important implication of the fundamental principle is that, according to GR, for a given experiment, all inertial observers must measure equal accelerations.  Mathematically, the equivalence of inertial frames is expressed by the basic Galilean coordinate transformations.  For a particle with position ${\bf x}_p$ in a frame $\cal S$ (say the block in Fig. \ref{block1}), its position in  frame $\cal S'$ moving at a velocity $\bf v$  with respect to $\cal S$ is
\be
{\bf x}_p' = {\bf x}_p -{\bf v}t. \label{Gal1}
\ee
Because $\cal S$ and $\cal S'$ are considered inertial, $\bf v$ is constant.  Thus, taking time derivatives, gives the Galilean law of velocity addition 
\be
{\bf v}_p'=  {\bf v}_p - {\bf v}, \label{Gal2}
\ee
with which any 19th century philosopher, changing trains, would agree.  Since $\bf v$ is constant,  taking a second time derivative gives
\be
{\bf a}'_p = {\bf a}_p, \label{Gal3}
\ee
establishing that  accelerations measured by Simplicio on a moving train are the same as those  measured by Salviati standing on a  station platform. When ${\bf F} = m {\bf a}$, any forces the two measure are therefore also identical.

Notice, however, in measuring ${\bf v}'_p$, Simplicio on the train is implicitly taking  $\frac{d{\bf x}_p'}{dt'} = \frac{d{\bf x}_p'}{dt}\frac{dt}{dt'}$, which  equals  $\frac{d{\bf x}_p'}{dt}$ only when $dt' = dt$.  The same holds for ${\bf a}'_p$.  Hence, in writing (\ref{Gal2}) and (\ref{Gal3}), we have tacitly assumed that the time intervals measured in both frames are the same:
\be
dt' = dt \qquad \mathrm{or} \qquad t' = t + \mathrm{constant}\label{Gal4},
\ee
which is the Galilean time transformation. No nineteenth-century natural philosopher would take issue with this assumption either.\\

\begin{figure}[htb]
\vbox{\hfil\scalebox{.6}
{\includegraphics{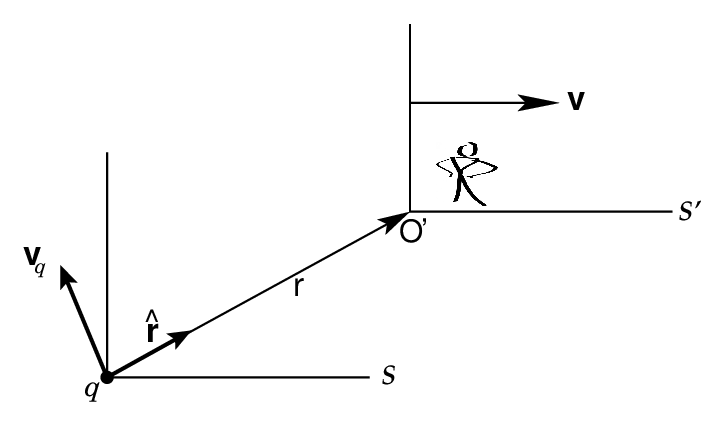}}\hfil}
{\caption{\footnotesize{A charge $q$ is located at the origin of $\cal S$ and moves with a velocity ${\bf v}_q$ in that frame.  The unit vector $\bf\hat r$ points from $q$ to the observation point $O'$, while $r$ is the distance between $q$ and $O'$.\label{charge1}}}}
\end{figure}

With (\ref{Gal2}) one can derive the Galilean transformations of the electric and magnetic fields, which can be found in both undergraduate and graduate textbooks (e.g. \cite{Knight07, Jackson99}). If a charge $q$ is moving at velocity ${\bf v}_q$ in $\cal S$, then by the Biot-Savart law its magnetic field $\bf B$ measured in $\cal S$ is, in Gaussian units,
\be
{\bf B} = \frac{q}{c} {\bf v}_q \times \frac{{\bf\hat r}}{r^2}, \label{Biot}
\ee
where $\bf\hat r$ is the unit vector pointing from $q$ to the observation point $O'$ and $r$ is the distance between them (Figure \ref{charge1}). Then  Galilean velocity addition (\ref{Gal2} implies
\be
{\bf B}' = \frac{q}{c}{\bf v}'_q \times \frac{{\bf\hat r}}{r^2} = \frac{q}{c}({\bf v}_q - {\bf v}) \times \frac{{\bf\hat r}}{r^2} = \left(\frac{q}{c} {\bf v}_q\times \frac{{\bf\hat r}}{r^2}\right)  - \left(\frac{q}{c}{\bf v} \times \frac{{\bf\hat r}}{r^2}\right) ,
\ee
or, noticing that  ${\bf E} = q {\bf\hat r}/{r^2}$ is the static Coulomb field produced by $q$ in $\cal S$,
\be
{\bf B}' = {\bf B} - \bf \bs\b \times E. \label{B'1}
\ee
This expression is exact, assuming Galilean transformations.

We can now find ${\bf E}'$ by demanding with Galileo that the forces on $q$ in $\cal S$ and $\cal S'$ be equal. The total force on $q$ is 
\be
{\bf F} = q({\bf E}  + \frac1{c}\bf  v \times B), \label{Lorforce}
\ee
in which case
\be
{\bf E}' + \frac{{\bf v}_q'}{c} \times {\bf B'} = {\bf E} + \frac{{\bf v}_q}{c} \times {\bf B}.\label{Lorforce2}
\ee
Substituting in (\ref{Gal2}) with  ${\bf v}_p' = {\bf v}_q'$ and (\ref{B'1}) for ${\bf B}'$ yields
\be
{\bf E}' = {\bf E} + \bs\b \times {\bf B} + {\cal O}(\b^2). \label{E'1}
\ee
Equations (\ref{B'1}) and (\ref{E'1}) can also be obtained by merely truncating the full relativistic transformations for $\bf E$ and $\bf B$ at first order (see (\ref{Etrans}-\ref{Btrans})).
\begin{flushleft}
{\it Salviati }: Permit me to interrupt.  Simplicio, I perceive that the force  you have just caused to be written down is nothing less than the famous Lorentz force law.  If the rules of this wager are  that   we restrict ourselves to physics   known to practitioners of the mid-nineteenth century, then surely you are violating your own terms. Forfeit the bottle.\\
\vspace{3mm}
{\it Simplicio  }: You are undoubtedly right, Salviati, that no natural philosopher of 1850 would have expressed it  this way, but the force law itself follows directly from Faraday's law of induction and I do not see that it expressly requires relativity.  For the present, I shall not concede.
\end{flushleft}

Simplicio is correct in that the Lorentz force law follows directly from Faraday's law of induction,  first published in 1832.\footnote{Faraday's law was discovered by Faraday in 1831, by Henry at the same time or even earlier, and published by Faraday  the following year.  The Lorentz force law began to assume its present form with Thomson's paper of 1881 \cite{Thom81}, which was corrected by Heaviside \cite{Heavi89}, and assumed its modern form with Lorentz in 1895 \cite{Whit17}.  To derive the Lorentz  law from Faraday's,  one writes the latter as $\oint ({\bf F}/q)\cdot d\bs\ell= -d/dt \int \bf {\bf B} \cdot d{\bf A}$, for line element $d\bs\ell$ and area element $d\bf A$. Employing the convective derivative $d/dt = \dpr/\dpr t + \bf v \cdot \nabla$, this becomes $\oint ({\bf F}/q)\cdot d\bs\ell = - \dpr/\dpr t \int {\bf B} \cdot d\bs A - \int ({\bf v} \cdot \nabla) {\bf B} \cdot d{\bf A}$. Recalling that $\nabla \cdot {\bf B} = 0$, one gets $\oint ({\bf F}/q)\cdot d\bs\ell = - \dpr/\dpr t \int {\bf B} \cdot d{\bf A} + \int \nabla \times ({\bf v \times B})\cdot d\bf A$.  Finally, with Faraday's law again and Stokes's theorem,  the right-hand side becomes $\oint( {\bf E +   v \times  B}) \cdot d\bs\ell$, giving the Lorentz force law (\ref{Lorforce}). For further details see \cite{Jackson99}.}   Salviati is surely correct in that no physicist wrapping coils in the mid-19th century would have considered the above argument, but the physics it contains was  available at the time.  
\begin{flushleft}
{\it Salviati }: I begin to perceive a faint grumbling from the spectators. \\
\vspace{3mm}
\end{flushleft}

Your ears are sharp, Salviati, but for the  moment let us placidly proceed with the implications of these transformations.  From Poynting's theorem \cite{Poynt84} (\S \;\ref{Ein06}), the momentum per unit volume of electromagnetic radiation is given by ${\bf p} = (1/{4\pi c})({\bf E \times B})$.  Also, the density of electromagnetic radiation is $u = (1/{8\pi}) (E^2+B^2)$, which seems to have been recognized by about 1860 \cite{Whit17}.  Notice that when $E = B$ (which we shall assume in the following), these two expressions immediately imply $p = { {\mc E}}/c$.  Most importantly for us, this is the case for plane electromagnetic waves. 

From field  transformations (\ref{B'1}) and (\ref{E'1}) we can calculate (${\bf E}' \times{\bf B}'$).  Assuming that $\cal S'$ is moving with velocity $+ v \bf\hat k$ with respect to $\cal S$,  the momentum density for the right-moving and left-moving beams becomes, respectively,
\be
{\bf p}'_{\g\pm} = \frac1{4\pi c} [\pm EB  - \b( E^2 + B^2) ]{\bf\hat k} + {\cal O}(\b^2).  \label{p+}
\ee
 Since $v' = -v$, (\ref{mom})  yields
\be
(m_i - m_f) v  =   -p'_{\g +} - p'_{\g -} = \frac{\b}{4\pi c} (E_+^2 + E_-^2 + B_+^2 + B_-^2) \equiv \frac{\b}{4\pi c} (E^2 +  B^2), \label{dp}
\ee
where it is understood that the quantities on the left are per unit volume.  

With $u \equiv (1/8\pi)({E}^2 + {B}^2)$ from above we have,  apparently, $\D mv = ({2v})/({c^2}) u$, 
which immediately implies  ${\cal E} = \frac12\D mc^2$.  
\begin{flushleft}
{\it Salviati }: Ah, Simplicio, you sly fox.  In the first place, you have assumed that the electromagnetic momentum is that given by Poynting's theorem, which  requires Maxwell's equations and which was not published until 1884.  Thus, not only could no natural philosopher in Faraday's wake have possibly produced it, but Maxwell's equations are relativistic and  so is this proof.  If that weren't enough, your answer is obviously incorrect.  You have therefore  shown yourself wrong on all counts.  Forfeit the bottle.  \\
\vspace{3mm}
{\it Simplicio  }: Hmm, if you permit me, Salviati, I believe the matter is not so simple.\\
\end{flushleft}

True.  Remembering to transform the energy density to $\cal S'$, one finds
\be
\begin{split}
u' &=\frac1{8\pi}({E'}^2 + {B'}^2)\\
&= \frac1{8\pi}\left[E^2 + B^2 + 2 {\bf E\cdot (\bs\b \times B)} -2 {\bf B\cdot (\bs\b \times E)}\right] + {\cal O}(\b^2).
\end{split}
\ee
 For $\bs\b = +\b\bf \hat k$ and $E = B$, the energy density for each beam in the rocket frame becomes simply
\be
u'_{\pm} = (1 \mp 2\b)u.  \label{u'}
\ee

This perhaps counterintuitive result, however,  is not the full story.   Imagine that the light is given off in $\cal S$ as a train of $n$ wavelengths $\l$ occupying a volume $V$. Then as measured in $\cal S'$ the energy contained in the same $n$ wavelengths is  ${\mc E}_\g' =  u'V' = u'V\l'/\l$.  With (\ref{u'})  and $\l'/\l = (1\pm\b)$ from the classical Doppler shift, 
\be
{\mc E}_{\g\pm}'= (1 \mp \b){\mc E}_{\g}. \label{Eg'}
\ee
The same considerations apply to the (total) momentum: 
\be
{p}_{\g\pm}'= (1 \mp \b){p}_{\g}. \label{pg'}
\ee

That is, when the volumes are properly transformed, we find that the energy and momentum in the beams are themselves classically Doppler shifted.\footnote{This is evidently the reason behind Einstein's observation \cite{Ein05a}  that ``it is noteworthy that the energy and the frequency of a light complex vary with the observer's state of motion according to the same law," although the demonstration presented here is the only nonrelativistic one we have seen.} This is extremely significant for today's debate because  (\ref{Eg'}) and (\ref{pg'}) show that the energy and momentum transform in the same way under a Galilean transformation.  Thus, since $p = {\mc E}/c$ is true for electromagnetic waves in the lab frame $\cal S$, it remains true in any frame ${\cal S}'$.  Consequently, the momentum-conservation equation (\ref{mom1}) for the thought experiment of \S 3 holds without change in the Galilean case and so does the result: ${\mc E} = \D mc^2$.\\

The surprises do not end there. In his 1905 experiment Einstein cleverly arranged for the two  bursts of radiation  to be emitted in opposite directions, ensuring that in  $\cal S$ the momentum of $m$ remains zero.  Algebraically, this corresponds to the cancellation of the first-order terms in $\b$ in (\ref{En'o}).  However, since these terms correspond to the momentum, let us be less clever than Einstein and consider only one burst of radiation emitted to, say, the left.  Thus (\ref{Eno}) becomes merely ${\cal E}_i = {\cal E}_f + {\cal E}_\g$ and  we retain only the first and last terms of (\ref{En'o}).  Momentum conservation in $\cal S$  now requires that $m$ recoils with momentum $m_fv_f = p_\g$.  A repetition of Einstein's argument gives 
\be
\mc E'_i - \mc E_i = \frac12 m_iv^2 \quad ;  \quad \mc E'_f - \mc E_f = \frac12 m_fv_f'^2  - \frac12 m_fv_f^2,
\ee
where this time  $v_f' = v-v_f$. Approximating the relativistic Doppler term in  (\ref{En'o}) as ${\cal E}_\g(1 + \b + \frac12\b^2)$   leads to
\be
\frac12(m_i - m_f)v^2 = ({\cal E}_\g\b - m_fv_fv) + \frac12{\cal E_\g}\b^2.
\ee
But as just mentioned $m_fv_f = p_\g$.  Equating like powers of $v$ immediately requires ${\cal E} = \D mc^2$ \emph{and} $p = {\cal E}/c$.

Evidently Einstein missed an opportunity  to kill two birds with one stone.  Salviati, however, with puckered lips, regards this result as fortuitous,  skeptical that if we retained the second-order terms in  Galilean  transformation \ref{E'1}),  the relationship $p = {\cal E}/c$ would be preserved.  

Remarkably it is.  With $\bf B' = B - \bs\b \times E$, but $\bf E' = E + \bs\b \times B - \bs\b\times(\bs\b \times E)$ from (\ref{Lorforce2}) $(v_q \ll v)$,  transforming  the energy and momentum densities gives to second order, after correcting for the volume contraction, 
\be
\mc E'_{\g\pm} = (1 \mp \b + \frac12 \b^2) \mc E_\g \quad ; \quad  p'_{\g\pm} = (1 \mp \b + \frac12 \b^2) p_\g, \label{E'p'vol}
\ee 
where again $p$  represents  the total momentum in the volume.   Einstein's  initial surprise  that energy transformed according to the relativistic Doppler shift is today redoubled: Beginning from rather asymmetric Galilean field transformations, the transformation factor in  (\ref{E'p'vol}) nevertheless remains, to second-order, the relativistic Doppler shift.    Again $\mc E = p/c$.

\begin{flushleft}
{\it Simplicio  }: So, you see, Salviati, in the end no one has mentioned photons. And while even the most brazen soul would not deny that   $p = \mc E/c$ is true under Einsteinian relativity,  it is also true under Galilean relativity.  Perhaps it is true under anybody's relativity.\\
\vspace{3mm}
{\it Salviati }: Hmm,  the conclusion is conceivably not entirely trivial.  Nevertheless, when all is said and done,  Simplicio, apart from employing  Newtonian  momentum and energy, the above demonstration merely illustrates  the compatibility of the  Doppler effect  with Maxwellian electrodynamics.  After all, the author has stubbornly persisted in assuming that light is transmitted by electromagnetic waves, whose existence was not even predicted before Maxwell, not to mention that the energy and momentum of these waves is given by Maxwell's equations, which are Lorentz invariant.  I do not see how you can get out of this.\\
\vspace{3mm}
{\it Simplicio  }: Well, umm, that is...\\ 
\end{flushleft}

What Salviati is attempting to say is that, although by the mid 19th century the wave picture of light was prevalent, the suggestion that light should be identified with \emph {electromagnetic} waves was published  by Maxwell only in 1865 \cite{Max65}.  Moreover, under a Lorentz transformation between  frames, Maxwell's equations remain unchanged, or as is often said, are ``Lorentz invariant,"\footnote {We prefer the term ``Lorentz covariant," meaning that Maxwell's equations retain their form in any inertial frame, although the values of the $\bf E$- and $\bf B$-fields have changed.}  which is precisely the reason Maxwell's electrodynamics are compatible with special relativity while  Newtonian physics is not.  

What Simplicio is attempting to say is that because we have employed only some approximation of Maxwell's equations, we  have manifestly broken their Lorentz invariance. Consequently, the extent to which ER lurks hidden in the above derivations remains ambiguous.  

One thing is clear: as the sun reaches the zenith we may no longer escape from attempting to make such notions more precise.

\section{Galileo Before the Tribunal}
\setcounter{equation}{0}\label{Gal}

By breaking Lorentz invariance, the Galilean field transformations (\ref{B'1}) and  (\ref{E'1}) have have left an opening for a possible non-ER derivation of ${\cal E} =mc^2$.  Alarmingly, however, fifty years ago Bellac and L\'evy-Leblond (BL)\cite{BLB73} argued that those  same ``undergraduate" transformations ``have no well-defined meaning and should be avoided altogether."  They then proposed rigorous Galilean limits to Maxwell's equations that are widely accepted today.  We must therefore justify our use of the undergraduate transformations before  audience grumbling becomes measurably louder.
\begin{flushleft}
{\it Salviati }:  That would  be prudent.\\
\end{flushleft}

Evidently, BL's principal objection to transformations (\ref{B'1}) and  (\ref{E'1}) was that they do not form a sensible group.   One should reasonably demand that transforming the electric or magnetic field first from frame  $\cal S$ to $\cal S'$,  moving at velocity $\bf v$ relative to $\cal S$, and then from $\cal S'$ to a frame $\cal S''$, moving at $\bf v$ relative to $\cal S'$, should yield the same result as  transforming directly from $\cal S$ to $S''$ with relative velocity $2\bf v$.  From (\ref{B'1}) and  (\ref{E'1}) it is simple to show that the two procedures do not  in fact agree.   Nevertheless, the results differ by ${\cal O} (\b^2)$, and so by today's rules we are entitled to say that the transformations behave sensibly  in the nineteenth century. 

The situation, however, is somewhat more complicated.  The limits proposed by BL do not coincide with either (\ref{B'1}) or  (\ref{E'1}) and so we must examine them in greater detail to determine whether they invalidate our deliberations.  BL term their ``strict" Galilean limits the ``electric limit" and the ``magnetic limit," and it is easiest to see where they come from by beginning with the full Lorentz transformation for the fields: 
 \bea
 {\bf E'} &=& \g\left[{\bf E} + \frac{\k_1}{\k_2 \a c}\bs\b \times {\bf B}\right] + \frac{(\g - 1)}{\b^2}{\bs \b}(\bs \b\cdot \bf E)\label{Etrans}\\
 {\bf B'} &=& \g\left[{\bf B} - \frac{\k_2 \a c}{\k_1}\bs\b \times {\bf E}\right] + \frac{(\g - 1)}{\b^2}{\bs \b}(\bs \b\cdot \bf B) \label{Btrans}
 \eea
 Here we follow the appendix of \cite{Jackson99} or \cite{Heras10} by writing the equations for arbitrary systems  of units. For all systems $\k_1/\k_2 = c_{units}^2$,  where $c_{u}^2 \equiv 1/{\m_0\e_0}$.  In the units used by most working physicists (Gaussian), $\k_1 = 4\pi, \k_2 = 4\pi/c_u^2$ and $\a = c$.   For Heaviside-Lorentz units $\k_1 = 1, \k_2 =1/c_u^2$ and $\a = c$; for SI units $\k_1= 1/\e_0, \k_2 =\m_0 $ and $\a = 1$. 

 We have done this because since  BL some physicists have objected to Gaussian units because they set $c = c_u$, where in the latter the permittivity of free space $\e_0$ and the permeability of free space $\m_0$ have no  {\it a priori} connection to $c$.  In fact, BL write, ``Any system of units, such as the CGS one, in which $c$ enters the very definition of the units is bound to give inconsistent results."
\vspace{3mm}
\begin{flushleft}
 {\it Simplicio  }: Umm, excuse me,  I have always been under the impression that physics must be independent of the system of units employed...\\
\vspace{3mm}
{\it Salviati }: With that every sensible person must agree.  I also fail to understand the meaning of Messieurs Bellac and L\'evy-Leblond's assertion.
\end{flushleft}

 In any case, BL were correct to point out that much confusion can arise because the speed of light appears in two guises in Maxwell's equations, the   first  the velocity of propagation, $c$,  the second  as the  constant $c_u$, which has dimensions of velocity.  Explicitly: 
\bea
\bs\nabla \cdot {\bf E} &= \k_1\r \qquad \quad &\mathrm{Gauss's \;law} \label{Gauss}\\
\bs\nabla \times {\bf E}+ \k_3\frac{\dpr \bf B}{\dpr t} &= 0 \qquad \quad \ \ \ &\mathrm{Faraday's \;law}\label{Faraday}\\
\bs\nabla \cdot {\bf B} &=0 \qquad \quad \ \ \ & \mathrm{Monopole \; exclusion} \label{GaussB}\\
\bs\nabla \times {\bf B}- \frac{\k_2\a}{\k_1} \frac{\dpr \bf E}{\dpr t} &= \k_2\a\bf J \qquad \quad &\mathrm{Amp\grave{e}re's \; law}\label{Ampere}
\eea
The constant $\a$ is freely specifiable for convenience (see \cite{Jackson99}).  By considering the electromagnetic wave equation, one finds that $\k_3\a = c_u^2/c^2$. If one sets $c = c_u$, then $\k_3$ in Faraday's law becomes identically $1/\a$. With the values from above, we see that in Gaussian units both $c$ and $c_u$ appear in Amp\`ere's law. 

None of this invalidates our deliberations to this point because by 1850 Fizeau  had measured $c$ to within five percent of the modern value, and in 1855 Weber and Kohlsrauch determined the value of $c_u$ to be about $3 \times 10^{10}$ cm/s \cite{Whit17}. In his famous 1857 article on the propagation of electricity in wires \cite{Kirch57}, Kirchoff  noted that $c_u$  is ``very nearly equal to the velocity of light in vacuo."  By 1862 Foucault had measured $c$ to within one percent of today's value.  Consequently, by then $c_u$ and $c$ might readily have been equated for experimental reasons if not theoretical ones.  Therefore we continue to use Gaussian units, seeing no harm and much convenience.

 Of greater importance, $\m_0$ and $\e_0$ are constants that  reflect the properties of the aether itself.  If one adopts the point of view of Preti et al. \cite{Preti09}, they are therefore {\it universal} constants, independent of reference frame.  In equating $c_u$ to the propagation speed of light, one must then accept that $c$ itself is a universal constant.  From this viewpoint, Simplicio had no necessity to suggest in \S 3 that the speed of light be transformed according to $c = c_o \pm v$, although, with certain  exceptions (\S 7) it is doubtful that many nineteenth-century  philosophers would have proposed otherwise.

Returning to the field transformations, we see that (\ref{B'1}) and  (\ref{E'1}) result directly  from  (\ref{Etrans}) and (\ref{Btrans}) by letting $\g \to 1$, which is merely a statement that $\b^2 \ll 1$ and not, contrary to what BL seem to imply, that $c\to \infty$.  Nonetheless,  one may then choose  either $E \gg  (\k_1/(\k_2 \a c)B$ or $B \gg  (\k_2 \a c/\k_1) E$. The former choice gives BL's ``electric limit":
\be
{\bf E}' = {\bf E} \ \quad ; \ \quad {\bf B}' = \bf B - \bs\b \times E, \label{Elim}
\ee
while the latter choice gives the ``magnetic limit":
\be
{\bf B}' = \bf B \ \quad ; \ \quad {\bf E}' = \bf E + \bs\b \times B. \label{Blim}
\ee
Again, neither of these corresponds to (\ref{B'1}) and  (\ref{E'1}). 
\vspace{3mm}
\begin{flushleft}
{\it Salviati }: Allow an interruption from me this time.  No  natural philosopher of the 19th century was aware of the Lorentz transformations.  This entire line of reasoning is a flagrant violation of the principle of chronological protection, and no rational person can believe that any physicist of the time could have imagined such limits. 
\end{flushleft}

Salviati has raised a cogent objection.  It is difficult to see what logic, working chronologically forward rather than backwards, would lead to these limits.  However, we are merely reporting the reasoning of BL and followers, who have  claimed to be pre-Maxwellians.

The limits do raise  concerns.  First, they are manifestly incapable of describing electromagnetic waves, for which $E=B$.  In fact, with the derivative operators
\be
\nabla' = \nabla \quad ; \quad \frac{\dpr}{\dpr t'} = \frac{\dpr}{\dpr t} + \bf v\cdot \nabla,
\ee
which follow from the Galilean coordinate transformations (\ref{Gal1}) and (\ref{Gal4}), one finds that not even the simplest Maxwell equation (\ref{GaussB})  transforms properly. 
One can also  confirm \cite{Preti09,Heras10}  that    Maxwell's equations in the electric and magnetic limits are

\hspace{3.1 cm} $Electric \  Limit$ \hspace{4.7 cm} $Magnetic \  Limit$
\begin{equation}
\begin{array}{rrrrrr}
\nabla \cdot {\bf E} &=& \k_1 \r &\hspace{3 cm}\nabla \cdot {\bf E} &=& \k_1 \r\nn\vspace{3mm}\\
\nabla \times {\bf E} &=& 0 &\hspace{ 3 cm} \nabla \times {\bf E} +\frac{\k_2}{\k_1 c^2}\frac{\dpr \bf B}{\dpr t} &=& 0 \nn\vspace{3mm}\\
\nabla \cdot {\bf B}  &=& 0 &\hspace{3 cm} \nabla \cdot {\bf B}  &=& 0\nn\vspace{3mm}\\
\nabla \times {\bf B} -\frac{\k_2}{\k_1} \frac{\dpr \bf E}{\dpr t} &=& \k_2 \bf J &\hspace{3 cm}
\nabla \times {\bf B} &=&\k_2 \bf J
\end{array}
\end{equation}

The usual procedure for deriving the electromagnetic wave equations is to take the time derivative of Faraday's law (\ref{Faraday}) and use Amp\`ere's law (\ref{Ampere}) without source to eliminate $\dot {\bf E}$, or vice versa.   Obviously that procedure  will not succeed here because in the electric limit the $\dot {\bf B}$ term is missing and in the magnetic limit the displacement current $\dot {\bf E}$ is missing.   Consequently, if one accepts the BL limits as the only valid ones to study Galilean electromagnetic waves, they cannot be studied.

An even more serious difficulty arises if we believe the fundamental principle of Galilean physics that measured accelerations must be the same in all inertial frames.  Requiring, as in (\ref{Lorforce2}), that the Lorentz force be equal in $\cal S$ and $\cal S'$,  velocity transformation (\ref{Gal2}) then implies 
\be
{\bf E}' + \frac{{\bf v}_q}{c} \times ({\bf B' - B}) = {\bf E} + \bs \b \times {\bf B'}.\label{Lorforce3}
\ee
Inserting (\ref{Elim}) into this equation  shows that the electric limit has no consistent solution for $\b \ne 0$.  Preti et al. \cite{Preti09}  argue further that the only physically sensible solution should be independent of ${\bf v}_q$, in which case the equation is consistent  with the magnetic limit only: ${\bf B}'=\bf B$ and ${\bf E}' = \bf E + \bs\b \times B$.  
\begin{flushleft}
 {\it Simplicio  }: I am much troubled by this discussion.  If Signori Bellac and L\'evy-Leblond are insisting on their limits, then they are elevating a mathematical expression above a physical principle. \\
\vspace{3mm}
{\it Salviati }: That appears to be the case.\\ 
\vspace{3mm}
{\it Simplicio}: Does that mean that mathematics is not always the language of the Book of Nature, contrary to what Galileo has written?
\end{flushleft}

What it does mean is that some non-Galilean coordinate transformations must be introduced in order to produce a set of ``Maxwell" equations that are valid through ${\cal O}(\b)$ and which admit electromagnetic waves. A first guess would be  the ``Quasi-Galilean" coordinate transformations
 \bea
t' = t - \frac{vx}{c^2} \quad &;& \quad  x' = x -vt\nn\\
t = t' +\frac{vx'}{c^2} \quad &;& \quad  x = x' +vt',  \label{QGal}
\eea
which are also ``Quasi-Lorentz" transformations ($\g = 1$).

\begin{figure}[htb]
\vbox{\hfil\scalebox{.6}
{\includegraphics{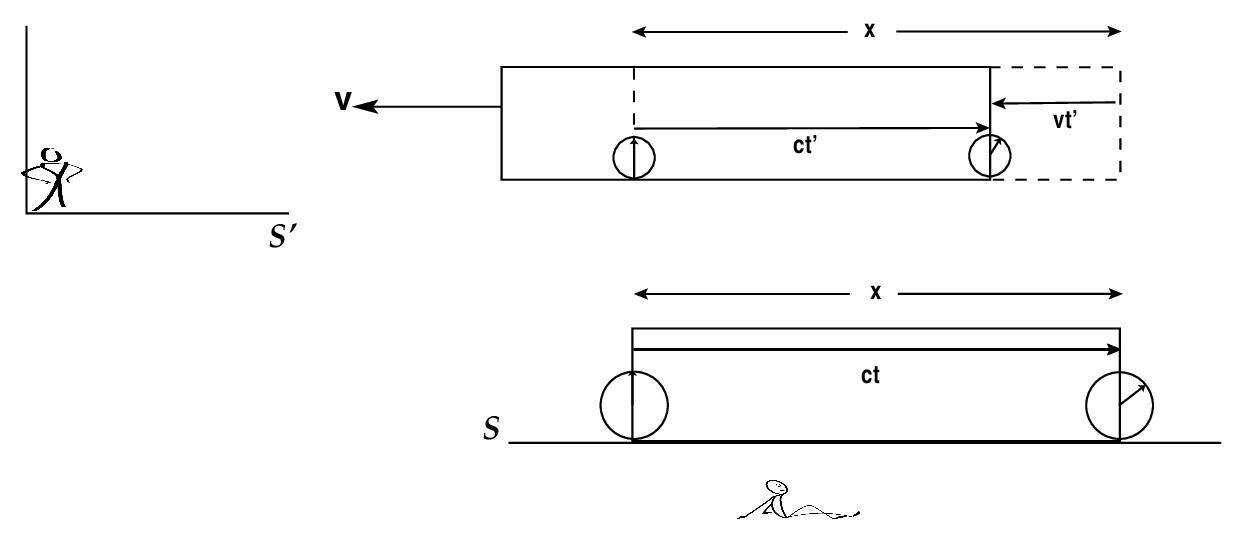}}\hfil}
{\caption{\footnotesize{Two synchronized clocks are located at the left and right ends of a lab of length $x$.  At $t=0$ a pulse of light is emitted to the right.  As observed in the lab frame $\cal S$ it hits the right wall at a time $t = x/c$.  In frame $\cal S'$ moving to the right at $v$, the light travels an approximate distance $ct' = x - vt$ before hitting the right wall, and so the light travel time is $t' = x/c - xv/c^2$, or $ t' = t - xv/c^2$ for an arbitrary distance $x$. \label{movcyl}}}}
\end{figure}

It is easy to convince oneself that this  system of  equations is not entirely self-consistent, but the inconsistencies appear at  ${\cal O}(\b^2)$. What do the transformations imply? Evidently, the $t$-transformation merely describes the amount that spatially separated clocks have become desynchronized due to the finite speed of light. As sketched in figure \ref{movcyl}, imagine two initially synchronized clocks separated by a distance $x$ in a lab.  In the lab frame $\cal S$, a light pulse emitted at $t=0$ by the clock on the left  will be received  the  clock on the right at a time $t = x/c$.   However, Simplicio in  frame $\cal S'$, moving at $\bf v$ with respect to the lab, will record that time as $t' = t -vx/c^2$. Therefore, these``Quasi-Galilean" transformations imply a relativity of simultaneity. However, time dilation and  Lorentz contraction, which require considering a round-trip of the light beam, are second-order corrections ($\g \ne 1$) and do not appear here. Hence, clocks must be measuring equal time intervals. Also, note that the entire discussion assumes that, as argued above, we accept that $c$ is a universal constant (or agree that transforming $c$ to $c \pm v$ will not change the result to first order).  

With coordinate transformations (\ref{QGal}) and the derivative operators
\be
\nabla_i' = \frac{\dpr t}{\dpr x^{i '}}\frac{\dpr}{\dpr t} + \frac{\dpr x^j}{\dpr x^{i '}}\frac{\dpr}{\dpr x^j} \quad ; \quad \frac{\dpr}{\dpr t'} = \frac{\dpr}{\dpr t} + \bf v\cdot \nabla,
\ee
one can  verify that  the field transformations  (\ref{B'1}) and  (\ref{E'1}) satisfy all  the standard Maxwell equations (\ref{Gauss})-(\ref{Ampere}) through ${\cal O}(\b)$.  In particular a displacement current is present.  They therefore admit wave solutions and one can talk about wave energy, momentum and Doppler shifts.

To summarize for the Tribunal, as the sun heads west we have  shown that the ``undergraduate'' field transformations, Quasi-Galilean coordinate transformations and Maxwell's equations  are all consistent with one another to ${\cal O} (\b)$. Because the ``Quasi-Maxwell" equations are {\it not} Lorentz invariant, we tentatively conclude that  the momentum-based proof of ${\cal E}=mc^2$ in \S 3 is valid without ER.  
\begin{flushleft}
{\it Salviati }: Hmm, I am more prepared to believe that L\'eon Foucault might have conjured up the ``Quasi-Galilean" transformations by considering the desynchonization of clocks due to light's finite speed than I am the ``electric" and ``magnetic limits," which require precognition of future centuries.  But only slightly.  I withhold comment on the general conclusion until the role of Poynting's theorem in the derivation is clarified.\\ 
\end{flushleft}

\section{Poynting's Theorem and Einstein's 1906 Derivation}
\setcounter{equation}{0}\label{Ein06}
\begin{flushleft}
{\it Salviati }:  At last.  Ah!\\
\end{flushleft}

Poynting's theorem is the work-energy theorem for the electromagnetic field, or an equation of continuity for the charge density (see, e.g., \cite{Jackson99}). In modern notation,  it reads
\be
{\bf J \cdot E} + \frac{\dpr{u}}{\dpr t} + \nabla \cdot \bf{ S} = 0, \label{Poynt}
\ee
where ${\bf J }\equiv \r_q \bf v$ is the current density for charge density $\r_q$; $u \equiv  \frac1{8\pi} (E^2 + B^2)$ and ${\bf S} \equiv \frac{c}{4\pi} (\bf E \times B)$.

This expression follows directly from Maxwell's equations with the help of standard vector identities. The first term represents work per unit time per unit volume performed by the electric field on the charges; thus one can identify $u$ with the energy density of the electromagnetic field, as we did in \S\ref{Galileo}.   The Poynting vector $\bf S$, which was also employed extensively in \S\ref{Galileo},  represents energy flux (energy/(area $\times$ time)) and when divided by $c^2$,  electromagnetic momentum density.

We argued in \S 5 that Maxwell's equations remain invariant under transformations \ref{QGal} to order ${\mc O}(\b)$. Therefore, (\ref{Poynt}) does as well.   This remark is relevant to one of Einstein's 1906 derivations of ${\mc E}=mc^2$ \cite{Ein06}, which relies directly on Poynting's theorem.  Not  so famous as the one of 1905, it is technically almost identical to Henri Poincar\'e's 1900 paper \cite{Poin00} in  which he argues that  the electromagnetic field may be regarded as a {\it fluide fictif} with  energy density such that ${\mc E}=mc^2$.  Einstein writes however: ``Although the simple formal considerations that have to be carried out...are in the main already contained in a work by H. Poincar\'e, for the sake of clarity I shall not base myself upon that work."  For the same reason the following should not be regarded as an exact transcription of Einstein's argument.

Both authors intend to show that assigning an effective mass to the electromagnetic field in an isolated  system consisting of radiation and charged masses requires the combined center of mass to move at constant velocity, as freshmen learn is required by momentum conservation.  Poincar\'e and Einstein imagine a stream of electromagnetic radiation impinging on a system of charged masses which are contained in some unspecified volume.  Einstein says at the outset that he is assuming that the kinetic energy of the masses is ``infinitesimally small" compared to their ``internal" energy, i.e., $(1/2)mv^2 \ll mc^2$.  This already signals he will ignore terms of order $v^2/c^2$.  He next multiplies (\ref{Poynt}) by a coordinate $x$,  representing particle position, and integrates over the volume of the system $\t$ to obtain
\be
\int x \, {\bf J \cdot E} \, d\t + \frac1{8\pi}\int x \frac{\dpr}{\dpr t}(E^2 + B^2)\,d\t
+ \int x \,\nabla \cdot {\bf S}\,d\t = 0. \label{Poynt2}
\ee
Taking $\bf S$ as a function of $x$ only and integrating the second and third terms by parts, (\ref{Poynt2}) can be rewritten as 
\be
\int x \, {\bf J \cdot E} \, d\t + \frac1{8\pi}\frac{\dpr}{\dpr t}\int x(E^2+B^2)\, d\t - \frac1{8\pi}\int (E^2+B^2)\frac{\dpr x}{\dpr t}\, d\t - \int { S}\, d\t = 0, \label{Poynt3}
\ee
where the boundary term $Sx$ has been discarded, evidently on the assumption that $\bf S$ vanishes at infinity.  Einstein  also omits the third term in (\ref{Poynt3}), apparently because he is taking particles velocities to be small, in which case he can write approximately  
\be
\int x \, {\bf J \cdot E} \, d\t + \frac1{8\pi}\frac{d}{dt}\int x(E^2+B^2)\, d\t - \int { S}\, d\t = 0. \label{Poynt4}
\ee
At this point Einstein does assign an equivalent mass density to the radiation: $\rho_{EM} \equiv \frac{(E^2 + B^2)}{8\pi c^2}$.  Furthermore, because $\bf J \cdot E$  represents the work per unit time performed by the electromagnetic field on a particle, Einstein  assumes that this results in a mass increase for each particle such that ${\bf J \cdot E} = \dot m c^2$, ignoring the kinetic energy.  He can then rewrite the first term to get
\be
c^2 \sum_i x_i \dot m_i + c^2\frac{d}{d t}\int x\, \r_{EM}\, d\t - \int { S}\, d\t = 0. \label{Poynt5}
\ee
Now, if $ \overline{\bf S} \equiv \frac1{c^2}\int {\bf S}\, d\t $ represents the total momentum of the radiation, then $d\overline{\bf S}/dt$ represents the total force exerted by the radiation  on the particles (which Einstein argues is the Lorentz force (\ref{Lorforce}).  Therefore,
\be
\frac{d}{dt}\left[\overline{ S } + \sum_i(m_iv_i)\right] = 0, 
\ee
or
\be 
\overline{ S}  + \sum_i(m_iv_i) = \mathrm{constant}. \label{Smv}
\ee
Further, the first term in (\ref{Poynt5}) can be rewritten as
\be
c^2 \sum_i x_i \dot m_i = c^2 \sum_i \frac{d (x_i m_i)}{dt} -c^2 \sum_i m_iv_i. \label{Poynt6}
\ee
Eliminating the final term in (\ref{Poynt6}) with (\ref{Smv}) and substituting the result back into (\ref{Poynt5}) leads to cancellation of the $\overline S$ terms.  So, finally,
\be
\frac{d}{dt} \left[\sum_i(x_im_i) + \int x\, \r_{EM}\, d\t\right] = \mathrm{constant}.
\ee
The bracketed expression represents the coordinates of the center of mass of the matter-radiation  and the equation tells us that it can move only  with a constant velocity. Thus, the assumption that ${\mc E}=mc^2$ is consistent with conservation of momentum when radiation is included in the system.
\begin{flushleft}
{\it Salviati }: Hmm, this demonstration is evidently open to many criticisms.\\ 
\vspace{3mm}
{\it Simplicio}: Alas, with that anyone would agree.\\  
\end{flushleft}

One is that Einstein has not clearly specified the system he is considering, that is, whether the particles and radiation are contained in a cavity of some sort.  If they are, he has neglected to include the cavity stresses (``Poincar\'e stresses") needed to contain the matter + radiation and which must be included in the energy balance.  

Most obviously, he is using the Newtonian expressions for  momentum and force, which means as forewarned he is ignoring terms of $\b^2$. Moreover, if by hypothesis the charged masses are absorbing radiation, their mass must increase. Thus the force on them should be calculated by the rocket equation $F = d (mv)/{dt} = m\dot v + \dot mv$, which differs from constant-mass case by the $\dot mv$ term. If ${\mc E} = mc^2$,  then  $\dot m v = {\dot { {\mc E}}v}/{c^2}$. But we  know $\dot{\cal E} \sim {\bf J \cdot E} \sim \r_q \bf v \cdot E$.  Hence $\dot m v\sim v^2/c^2$  and, although we did not explicitly take this into account above, in discarding terms of ${\cal O} (\b^2)$, Einstein is ignoring the distinction between $F = ma$ and $F = dp/dt$.\footnote{The meaning of $m$ in the various thought experiments is not always transparent.  Because Einstein's 1905 derivation assumes Newtonian kinetic energy,  $m$ evidently represents the rest mass. Similarly for  Rohrlich's version \cite{Rohr90} of the \S 3  experiment.  Steck and Rioux \cite{SR83} state  that their  version ``suggests"   $\mc{E} = mc^2$ holds for both relativistic and rest mass.  Above, since Einstein is ignoring terms of $\mc{O}(\b^2)$, the difference between the two becomes immaterial.  For a discussion of such issues see Okun \cite{Okun89}.}
\begin{flushleft}
{\it Salviati }: It would appear that this demonstration is even less convincing than the previous ones, not to mention that you would be hard pressed to name it ``perhaps-Maxwellian."\\
\vspace{3mm}
{\it Simplicio}: You may be correct, Salviati.  But if you accept that  Poynting's theorem holds at the same level of approximation that we have been considering, then I would say this proof, which proceeds from somewhat different assumptions---and which  does not  even require the Doppler shift, $E=B$, electromagnetic waves or apparently $p = {\cal E}/c$---is compatible with the others of first order and thus does not contain Einsteinian relativity.\\ 
\end{flushleft}

\section{As the sun sets...}
\setcounter{equation}{0}\label{Other}

There are other theories that conceivably allow derivations of ${\mc E}=mc^2$ in nonrelativistic approximations.  For instance, independently of Maxwell, in 1867 Danish physicist Ludwig Lorenz (1829-1891) published a theory of light \cite{Lorenz67,Kragh18} that was  partly based on Kirchoff's earlier work \cite{Kirch57} concerning the propagation of electrical currents along telegraph wires.  In many respects Lorenz's theory is pertinent to today's debate.  First,  he did not conceive of light as an electromagnetic wave, but rather as an electrical oscillation propagating through a conducting ``vacuum" of space (perhaps analogous to a plasma wave propagating through an ionized medium).  Nevertheless, he contended that the velocity of light was a constant, $c$, and moreover rejected the aether.  As he wrote at the close of his paper, ``In this idea there is scarcely any reason for adhering to the hypothesis of an aether...[T]he result of the present investigation, that the vibrations of light are electrical currents, has been obtained without the assumption of a physical hypothesis,and will therefore be independent of one." 

That in itself allows us to dispense with the complication of a variable speed of light, which has  periodically troubled Simplicio throughout the day.  Closely related to a constant $c$ was Lorenz's major contribution: the introduction of retarded potentials. Because light propagated at finite velocity, Lorenz found it reasonable that electrical disturbances measured at the present moment depended on the state of the system at the earlier time $t_r = t - r/c$, where $r \equiv |{\bf x} - {\bf x}'|$ is the distance between the source point $\bf x'$ and  observation point $\bf x$.  He was thus led to introduce the retarded scalar and vector potentials 
\be
\Phi({\bf x}) = \int \frac{\r({\bf x}', t_r)}{r}d\t' \quad ; \quad {\bf A (x)} = 
\frac1{c}\int \frac{{\bf J} ({\bf x}', t_r)} {r} d\t' \label{retpot}
\ee
in use even today. With the equation of continuity for charge density $\r$, these definitions satisfy a physical condition that was for over a century mistakenly called the ``Lorentz gauge," the requirement that $\nabla \cdot {\bf A} = -(1/c)d\Phi/dt$.  Then, by  taking $\nabla^2 \F, \nabla^2\bf A$, which depend on the retarded time, one finds that they  produce  the same inhomogeneous wave equations  usually derived by expressing  Maxwell's equations in terms of the nonretarded potentials $\F ({\bf x},t), {\bf A} ({\bf x}, t)$ and imposing the same condition.   Lorenz, though, accomplishes this without any  displacement current, which is universally said to ensure the Lorentz invariance of Maxwell's theory and allow electromagnetic waves. 

Lorenz's wave equations are manifestly Lorentz invariant as well, however, and the predictions of his theory for light are exactly those of Maxwell's \cite{Wong10}. Consequently, any derivations of ${\mc E}=mc^2$ involving light in  Lorenz's theory must be subject to the same considerations and approximations already discussed. 

In the 1860s none of this was established, but the recognition of vector potentials makes it possible to imagine a ``Victorian" demonstration of ${\mc E}=mc^2$ that might have been proposed by Lorenz or Maxwell even before the identification of light with electromagnetic waves.  The impetus for this experiment can be found in Maxwell's own writings. Noticing the resemblance between ${\bf F} = d{\bf p}/dt$ and ${\bf E} =-(1/c)d{\bf A}/dt$, Maxwell postulated that the quantity $q{\bf A}/c$ could be identified with ``electromagnetic," or ``potential," momentum. (Admittedly, he was cautious enough to allow that ``all such phrases...are to be considered as illustrative, not as explanatory"  \cite{Max65, Max73, JaOk01}.)  Nevertheless, the view that $\bf A$ represents a genuine momentum per unit charge was apparently prominent at the time \cite{ST96}. The association can be put on a firmer footing by writing the Lorentz force law (\ref{Lorforce}) in terms of the potentials instead of the fields.  For the force on a test charge $q_t$, one finds
\be
{\bf F} \equiv  \frac{d{\bf p}_{mech}}{dt} = -q_t\nabla[\Phi -\frac1c({\bf v\cdot A})] -\frac{q_t}{c}\frac{d{\bf A}}{dt},
\ee
which prompts us to  identify   ``potential momentum" ${\bf p}_A$ with $q_t{\bf A}/c$.\footnote{See \cite{ST96, Kon78} for several arguments and experiments that strengthen the identification. The quantity  ${\bf p = p}_{mech} + q_t{\bf A}/c$ is of course the generalized, or canonical, momentum.}

Let us  now take Maxwell at his word and find ${\bf p}_A$ produced by a single  source charge $q_s$ moving at velocity ${\bf v}_s$.  Eq. (\ref{retpot})  for the nonretarded  case gives ${\bf A} = ({q_s}{\bf v}_s)/({c}{r})$, implying that
\be
{\bf p}_A = \frac{q_sq_t}{c^2r} {\bf v}_s. \label{pA}
\ee

To explore mass-energy conversion Queen Victoria might now  propose an experimental arrangement similar to that in \S 3, except that this time we consider two identical masses $m$ separated by a distance $r$,  each carrying  charge $q$ and moving toward each other in frame $\cal S$ at constant velocities $\pm v_q$. (Victoria assumes  large enough masses to make accelerations due to Coulomb forces negligible.) In one dimension, each charge produces a vector potential at the other $A = \pm({q}/{cr}){v}_{q}$.\footnote{It is perhaps easiest to visualize these as retarded potentials that are switched on at a time $t = 0$ in the past and propagate toward the other charge; however the use of potential momenta does not require they be retarded.   In fact, since  ${\bf A}_{r} = {\bf A}(t_r=t-r/c)$, then  ${\bf A}_{r} \approx {\bf A}_t + \dot {\bf A}|_t \D t = {\bf A}_t -r/c \dot {\bf A}|_t = {\bf A}_t (1+v/c)$.  Thus any corrections introduced by retarded potentials to the experimental result will be of order $\b^2$.} 

Taking potential momentum seriously, in $\cal S$ the momentum of each mass is $p_\pm = \pm mv_q \mp (q^2/c^2r)v_q$ and the total momentum of the system $p_+ + p_- = 0$.  In $\cal S'$, moving relative to $\cal S$ with velocity $+v$,
\be
p'_{+} = m(v_q - v) -\frac{q^2}{c^2 r}(v_q + v) \quad ; \quad p'_{-} = -m(v_q+v) +\frac{q^2}{c^2 r}(v_q -v). 
\ee
Thus, $p'_{tot} = p'_+ + p'_- = -2mv -2q^2v/(c^2r)$.

If we fervently believe that momentum should be conserved, then without even a physical collision among the masses we must have 
\be
(\Delta m)v = -\Delta \left(\frac{q^2}{r}\right)\frac{v}{c^2}.
\ee
Victoria, though, quickly recognizes the quantity in parentheses as the energy of $m$ in the Coulomb field and concludes that from this expression flows the unexpected result $\Delta m c^2 = -\Delta {\cal E}$, indicating that as the energy of the field decreases, the total mass increases.  She notices no mention  of electromagnetic waves or $p = {\cal E}/c$. 
\begin{flushleft}
{\it Simplicio}: Hmm, I do find this demonstration fairly...somewhat...nearly...convincing.\\
\vspace{3mm}
{\it Salviati }: Not I.  Apart from any other ambiguities, the author has  assumed that the energy is the Coulomb energy, neglecting the  magnetic energy, and has employed the same Newtonian expression for momentum as before.\\
\vspace{3mm}
{\it Author}: Both assumptions should be valid for our agreed-upon approximations.\\
\vspace{3mm}
{\it Simplicio}: Yes, and I would therefore  say that it is no better or worse than the other derivations.\\
\vspace{3mm}
{\it Salviati }: But my dear Simplicio, that is precisely the point, and I therefore say you must forfeit the bottle, for after an entire day of discussion you have failed to produce a nonrelativistic derivation of ${\cal E}=mc^2$ that is free of  inconsistencies and ambiguities.  What is more, given the basic dimensions involved, no one can rule out a coincidence---virtually any proof employing such elementary considerations may well give the same answer.\\ 
\vspace{3mm} 
 {\it Simplicio}: But my dear Salviati, I declare that is  the point.  The demonstrations we have seen are certainly inconsistent, but neither more nor less inconsistent than Einstein's own.   In some respects I do admit to finding  myself more confused then when we began.  Still, I shall not forfeit the bottle, for while it is surely the case that ${\cal E}=mc^2$ is an experimentally confirmed truth,  the truth of the derivations---well, that is a swamp of a different color... \\ 
\vspace{3mm} 
{\it Salviati }: By chance, have you heard the rumor that Amp\`ere  had nothing to do with Amp\`ere's circuital law?\\ 
\vspace{3mm}
{\it Simplicio}: You cannot be serious. \\ 
\vspace{3mm}
{\it Author}: That  appears to be true \cite{Erlichson99}. \\
\vspace{3mm}
{\it Simplicio}: Then where, pray tell, does that leave us? \\ 
\vspace{3mm}
{\it Salviati }: I suggest we open the bottle. Surely answers shall  flow forth...\\
\vspace{3mm}
{\it Simplicio}:  An agreeable proposition...Allow me to confess  another matter that has been troubling me...
\end{flushleft}

\vspace{2 cm}

\flushleft {\bf Acknowledgments}:

My gratitude to Stephen Boughn, for several critical readings of the manuscript,  numerous helpful and substantive suggestions, and above all for neverending debates, past, present and future.  Thanks also to the referees for their positive suggestions.\\

{\bf Request}:
Authors are requested to submit their personal proofs of $E=mc^2$ to an appropriate journal.

{\small

\begin{thebibliography}{10}

\bibitem{BLB73}
M.~Le Bellac and J.-M.~L\'evy-Leblond, ``Galilean Relativity," \emph{Il Nuovo Cimento}, {\bf 14}B, N. 2, 217--233 (1973).

\bibitem{Ein05b}
 A.~Einstein, ``Ist die Tr\"agheit eines K\"orpers von seinem Energieinhalt abh\"angig?"  \emph{Annalen der Physik} {\bf 323 (18)}, 639–-641 (1905).  English translation as ``Does the inertia of a body depend on its energy content?" in \emph{Einstein's Miraculous Year} (Princeton University Press: Princeton, 1998).

 \bibitem{Thom81}

J.~Thomson, ``On the electric and magnetic effects produced by the motion of electrified bodies," \emph{Phil. Mag.}, Ser. 5, {\bf 11}, 229--249 (1881).

\bibitem{Heavi89}

O.~Heaviside, ``On the electromagnetic effects due to the motion of electrification through a dielectric," \emph{Phil. Mag.}, Ser. 5, {\bf 27}, 324--339 (1889).

\bibitem{Ab03}

M.~Abraham, ``Prinzipien der Dynamik des Elektrons," \emph{Annalen der Physik} {\bf 315 (1)}, 105--179 (1903).

\bibitem{Poin00}

H.~Poincar\'e, ``La th\'eorie de Lorentz et le principe de r\'eaction," \emph{Archives néerlandaises des sciences exactes et naturelles} {\bf 5}, 252–-278 (1900).

\bibitem{H1}
F.~Hasen\"ohrl, ``Zur Theorie der Strahlung in bewegten K\"orpern," \emph{Wiener Sitzungsberichte} {\bf 113}, 1039--1055 (1904).

\bibitem{H2}
F.~Hasen\"ohrl, ``Zur Theorie der Strahlung in bewegten K\"orpern," \emph{Annalen der Physik} {\bf 320}, 344--370  (1904).

\bibitem{H3}
F.~Hasen\"ohrl, ``Zur Theorie der Strahlung in bewegten K\"orpern, Berichtigung," \emph{Annalen der Physik} {\bf 321}, 589--592 (1905).

 \bibitem{Jammer51}
M.~Jammer, \emph{Concepts of Mass} (Dover: Mineola, 1961).

\bibitem{Fadner88}
W.~Fadner, ``Did Einstein really discover $E=mc^2$?" \emph{Am. J. Phys.} {\bf 56}, 114–-122 (1988).

\bibitem{Roth21a}
T.~Rothman, ``The Curse of $E=mc^2$," \emph{American Scientist}, {\bf 109}, Nov.-Dec., 360--367 (2021). 

\bibitem{BR11}
S.~Boughn and T.~ Rothman, ``Hasen\"ohrl and the equivalence of mass and energy," arXiv:1108.2250v4.

\bibitem{Roth21}
T.~Rothman, ``Addendum to Hasen\"ohrl and the equivalence of mass and energy," arXiv:2105.09997.

\bibitem{Boughn13}
S.~Boughn, ``Fritz Hasenohrl and $E = mc^2$," \emph{ Eur. Phys. J. H.} {\bf 38}, 262--78 (2013).

\bibitem{Laue11}
 M.~von Laue, \emph{Das Relativit\"atsprinzip} (Vieweg: Braunschweig, 1911).

\bibitem{Klein18}
 F.~Klein, ``\"Uber die Integralform der Erhaltungss\"atze und der Theorie die r\"aumlich-geschlossenen Welt," \emph{Nach. Gesell. Wissensch. G\"ottingen, Math.-Physik, Klasse}, 394--423 (1918).

\bibitem{Ohan12}
H.~Ohanian,``Klein's theorem and the proof of $E=mc^2$," \emph{Am. J. Phys.} {\bf 80}, 1067–-072 (2012). 

 \bibitem{Ohan09}
 H.~Ohanian, ``Did Einstein Prove $E=mc^2$?" \emph{Studies in History and Philosophy of Modern Physics} {\bf 40}, 167--173 (2009).

\bibitem{Ein05a}
 A.~Einstein, ``Elektrodynamik bewegter K\"orper,"  \emph{Annalen der Physik} {\bf 322 (17)}, 891-–921 (1905). English translation as ``On the electrodynamics of moving bodies" in \emph{Einstein's Miraculous Year} (Princeton University Press: Princeton, 1998).

 \bibitem{Planck07}
 M.~Planck,``Zur Dynamik bewegter Systeme," \emph{Sitz. der Preuss. Akad. Wiss.}, June 13, 542-570 (1907).
 English translation as ``On the dynamics of moving systems," https://en.wikisource.org/wiki/Translation:On\_the\_Dynamics\_of\_Moving\_Systems.

 \bibitem{Mink08}
 H.~Minkowski, ``Space and Time," address delivered to the 80th Assembly of German Natural Scientists and Physicians, English translation reprinted in \emph{The Principle of Relativity} (Dover Publications: NY, 1952).

 \bibitem{SR83}
 D.~Steck and F.~Rioux, ``An elementary development of mass-energy equivalance," \emph{Am. J. Phys.} {\bf 51}, 461--462 (1983).

 \bibitem{Rohr90}
 F.~Rohrlich, ``An elementary derivation of $E=mc^2$," \emph{Am. J. Phys.} {\bf 58}, 348--349 (1990).


\bibitem{Compton23}
A.~ Compton, ``A Quantum Theory of the Scattering of X-Rays by Light Elements," \emph {Phys. Rev.} {\bf 21}, 483–-502 (1923).

\bibitem{Ein16}
A.~Einstein,``Zur Quantentheorie der Strahlung," \emph {Mitteilungen der Physikalischen Gesellschaft zu Zürich} {\bf 16} 47.   English translation as ``On the Quantum Theory of Radiation," in \emph{The Collected Papers of Albert Einstein} {\bf 2}, The Berlin Years, 1914-1917, Doc. 38, 220-233 (Princeton University Press: Princeton, 1996). 

\bibitem{Pesk10}
C.~Peskin, ``Wave momentum," \emph{The Silver Dialogues in Arts and Science} (New York University, 2010).

\bibitem{RP99}

D.~Rowland and C.~Pask, ``The missing wave momentum mystery," \emph{Am. J. Phys.} {\bf 67}, 378--388 (1999). 

\bibitem{Knight07}
R.~Knight, \emph{ Physics for Scientists and Engineers, A Strategic Approach}, second edition (Pearson-Addison Wesley: San Fransisco, 2007).

\bibitem{Jackson99}
J.~Jackson, \emph{Classical Electrodynamics}, third edition (John Wiley and Sons: New York, 1999).

\bibitem{Whit17}
E. Whittaker, \emph{ A History of the Theories of Aether and Electricity}, (Dover: Mineola, 2017) vol. I, chap. 13.

\bibitem{Poynt84}
J.~Poynting, ``On the Transfer of Energy in the Electromagnetic Field," \emph{ Phil. Trans.   Roy. Soc. Lon.} {\bf 175}, 343--361 (1884).

\bibitem{Max65}
J.~Maxwell, ``A Dynamical Theory of the Electromagnetic Field," \emph{Phil. Trans. Roy. Soc.} {\bf 155}, 459--512 (1865), in particular Arts. 20, 72.  

\bibitem{Heras10}
 J.~ Heras, ``The Galilean limits of Maxwell's equations," \emph{Am. J. Phys.} {\bf 78}, 1048--1055  (2010).
 
\bibitem{Kirch57}
G.~Kirchhoff, ``On the motion of electricity
in wires," \emph{Phil. Mag.} Ser. 4, 393--412 (1857).

\bibitem{Preti09}
G.~Preti et al., ``On the Galilean non-invariance of classical electromagnetism," \emph{Eur. J. Phys.} {\bf 30}, 381–-391  (2009).  

 \bibitem{Ein06}
A.~Einstein, ``Das Prinzip von der Erhaltun des Schwerpunktsbewegun un die Tr\"agheit der Energie,"\emph{Annalen der Physik} {\bf 20}, 627--633 (1906).  English translation as ``The Principle of Conservation of Motion of the Center of Mass of Gravity and the Inertia of Energy" in \emph{The Collected Papers of Albert Einstein} {\bf 2}, The Swiss Years, 1900-1909, Doc. 35, 200-206 (Princeton University Press: Princeton, 1989). 

\bibitem{Okun89}
L.~Okun, ``The Concept of Mass," \emph{Physics Today}, June, 31--36 (1989).

\bibitem{Lorenz67}
L.~Lorenz, ``On the identity of the vibrations of light with electrical currents," \emph{Phil. Mag.} Ser. 4 {\bf 34}, 287--301 (1867).

\bibitem{Kragh18}
H.~Kragh, ``Ludvig Lorenz and His Non-Maxwellian Electrical Theory of Light," \emph{Phys. Perspect.} (2018),
https://doi.org/10.1007/s00016-018-0223-1.

\bibitem{Wong10}
C.~Wong, ``Lorenz’s electromagnetic theory of light," arXiv: 1012.412v1.

\bibitem{Max73}
J.~Maxwell, ``A Treatise on Electricity and Magnetism," {\bf II} Art. 618, p. 236 (Clarendon Press: Oxford, 1873).

\bibitem{JaOk01}
J.~Jackson and L.~Okun, ``Historical roots of gauge invariance," \emph{ Rev. Mod. Phys.} {\bf 73}, 663--680 (2001).

\bibitem{ST96}
M.~Semon and J.~Taylor, ``Thoughts on the magnetic vector potential," \emph{Am. J. Phys.} {\bf 64}, 1361–-1369 (1996). 

\bibitem{Kon78}
E.~Konopinski, ``What the electromagnetic vector potential describes," \emph{Am. J. Phys.} {\bf 46}, 499–-502 (1978). 

\bibitem{Erlichson99}
H.~Erlichson, ``Ampère was not the author of `Ampère’s Circuital Law,'" \emph{Am. J. Phys.} {\bf 67}, 448–-450 (1999).
\end{thebibliography}
\end{document}